\documentclass[twocolumn,showpacs,preprintnumbers,amsmath,amssymb]{revtex4}

\usepackage{graphicx}
\usepackage{dcolumn}
\usepackage{bm}

\begin{document}

\preprint{AP-06}

\title{Cr-doping effect on the orbital fluctuation of \\
heavily doped Nd$_{1-x}$Sr$_x$MnO$_{3}$ ($x$ $\approx$ 0.625)}

\author{R. Tasaki}
\email{r-tasaki@sophia.ac.jp}

\author{S. Fukushima}
\author{M. Akaki}
\author{D. Akahoshi}
\author{H. Kuwahara}
\affiliation{
Department of Physics, Sophia University\\
Chiyoda-ku, Tokyo 102-8554, JAPAN
}%

\date{\today}


\begin{abstract}
We have investigated the Cr-doping effect of Nd$_{0.375}$Sr$_{0.625}$MnO$_3$ near the phase boundary between the $x^2-y^2$ and $3z^2-r^2$ orbital ordered states, where a ferromagnetic correlation and concomitant large magnetoresistance are observed owing to orbital fluctuation. Cr-doping steeply suppresses the ferromagnetic correlation and magnetoresistance in Nd$_{0.375}$Sr$_{0.625}$Mn$_{1-y}$Cr$_y$O$_3$ with $0 \leq y \leq 0.05$, while they reappear in $0.05 < y \leq 0.10$. Such a reentrant behavior implies that a phase boundary is located at $y = 0.05$, or a phase crossover occurs across $y = 0.05$.

\end{abstract}

\maketitle

%

\subsection{\label{sec:Introduction}Introduction}
 Mn oxides with a perovskite structure have attracted much attention because of the colossal magnetoresistance (CMR) effect\cite{Dagotto_PR_344,Tokura_RPP_69}. Since the magnetic and transport properties of the perovskite manganites are strongly affected by ordering patterns of $x^2- y^2$ and/or $3z^2 - r^2$ orbitals, a detailed investigation of the orbital-ordered (OO) states is significant for understanding the CMR effect. In heavily doped Nd$_{1-x}$Sr$_x$MnO$_{3}$ (NSMO), there exist two types of OO states, which exhibit highly anisotropic magnetic and transport properties\cite{Kajimoto_PRB_60}. One is the $x^2-y^2$ OO state ($0.53 \leq x < 0.63$), accompanying the $A$-type antiferromagnetic (AF) order in which the $x^2-y^2$ electrons are conducting within the ferromagnetic (F) plane\cite{Kuwahara_PRB_82}. The other is the $3z^2-r^2$ OO state ($0.63 \leq x \leq 0.80$), accompanying the $C$-type AF order. These two OO states compete with each other in a bicritical manner at $x = 0.625$\cite{Kajimoto_PRB_60}. Near the bicritical region, competition between the two OO states causes spatial orbital fluctuation on nanometer scale, which gives rise to the F correlation and concomitant large magnetoresistance (MR)\cite{Akahoshi_PRB_77,Nagao_JPCM_19}.
 
 It is well-known that the presence of quenched disorder in a bicritical (or multicritical) region where a ferromagnetic metallic (FM) and AF insulating states meet often causes phase separation phenomena, which are essential for the CMR. In Nd$_{0.5}$Ca$_{0.5}$MnO$_3$, for example, Cr-substitution on Mn-sites turns the charge- and orbital-ordered (CO/OO) state into the FM one\cite{Kimura_PRB_62,Kimura_PRL_83}.

 Therefore, it can be expected that Cr-doping into NSMO near the bicritical region ($x = 0.625$) induces phase separation and/or enhances the orbital fluctuation, which might lead to nontrivial phenomena such as the CMR. In this study, we have investigated the Cr-doping effect of Nd$_{0.375}$Sr$_{0.625}$Mn$_{1-y}$Cr$_y$O$_3$(NSMCO) ($0 \leq y \leq 0.10$).

%
%
\subsection{\label{sec:Experiment}Experiment}
 NSMCO crystals with $0 \leq y \leq 0.10$ were prepared using the floating zone method. We confirmed that all synthesized crystals are of single phase by the powder X-ray diffraction method. Magnetic and transport properties were measured using a Quantum Design physical property measurement system (PPMS). We randomly cut the synthesized crystals with the size larger than twin-domain size for measurements of magnetic and transport properties.

%
%
\subsection{\label{sec:Results and discussion}Results and discussion} 
\begin{figure}
\includegraphics[clip]{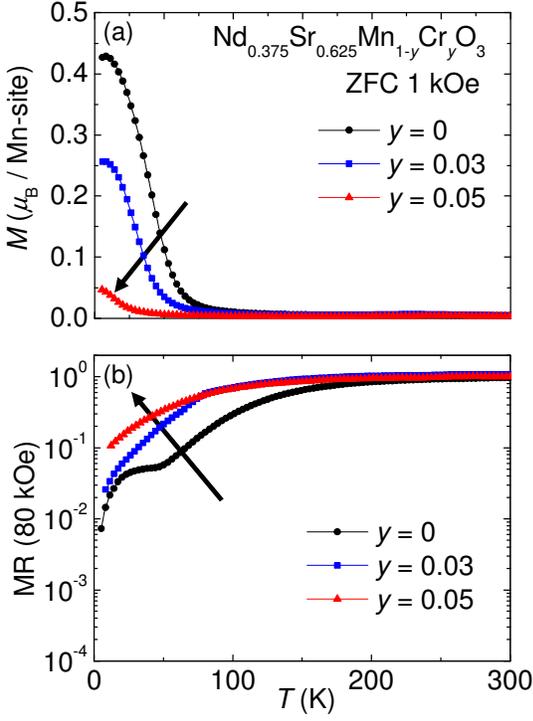}
\caption{\label{fig:fig1} (Color online) Temperature ($T$) dependence of (a) magnetization ($M$) and (b) magnetoresistance [MR(80 kOe)] of Nd$_{0.375}$Sr$_{0.625}$Mn$_{1-y}$Cr$_y$O$_3$(NSMCO) with $y$ = 0, 0.03, and 0.05. ZFC represents zero field cooling process. MR(80 kOe) is defined as MR(80 kOe) $\equiv \rho$(80 kOe) / $\rho$(0 Oe).}
\end{figure}

 First, we show in Figs. 1(a) and 1(b) the temperature ($T$) dependence of the magnetization ($M$) and MR(80 kOe) of NSMCO with $0 \leq y \leq 0.05$, respectively. Here MR(80 kOe) is defined as MR(80 kOe) $\equiv \rho$(80 kOe) / $\rho$(0 Oe), where $\rho$(0 Oe) and $\rho$(80 kOe) are resistivities measured in $H$ = 0 Oe and 80 kOe, respectively. In $y$ = 0, the F correlation is observed due to the orbital fluctuation below 65 K, and the MR(80 kOe) at 5 K is below 0.01: the resistivity drops more than two orders of magnitude by applying a magnetic field of $H$ = 80 kOe. This result is consistent with our previous report\cite{Akahoshi_PRB_77}. Cr-doping steeply suppresses the F correlation and concomitant MR, which are most suppressed at $y$ = 0.05. 
 
\begin{figure}
\includegraphics[clip]{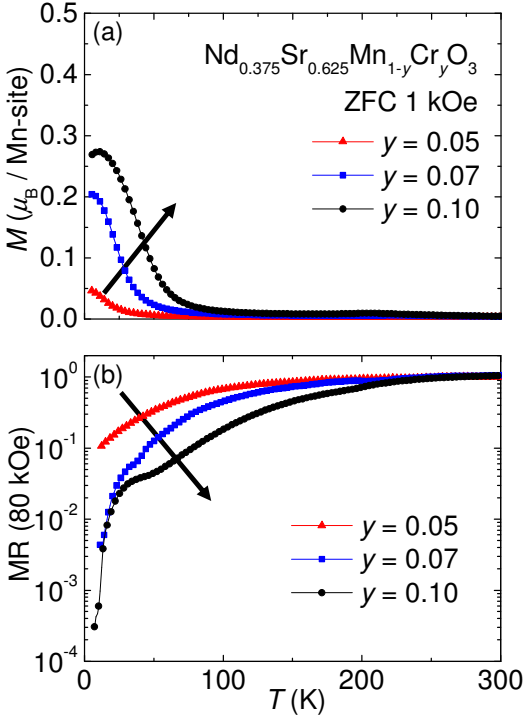}
\caption{\label{fig:fig2} (Color online) $T$ dependence of (a) $M$ and (b) MR(80 kOe) of NSMCO with $y$= 0.05, 0.07, and 0.10.}
\end{figure}
Then, we exhibit the $T$ dependence of the $M$ and MR(80 kOe) of NSMCO with $0.05 \leq y \leq 0.10$ in Figs. 2(a) and 2(b). In $y > 0.05$, the F correlation and MR reappear and are evolving with an increase of $y$ from 0.05 to 0.10. Note that the $M$ and MR of $y = 0.10$ are quite similar to those of $y = 0$, as clearly seen from Figs. 1 and 2, indicating that the F correlation and MR of NSMCO show a reentrant behavior with a change of Cr-concentration $y$.
 
 \begin{figure}
 \includegraphics[clip]{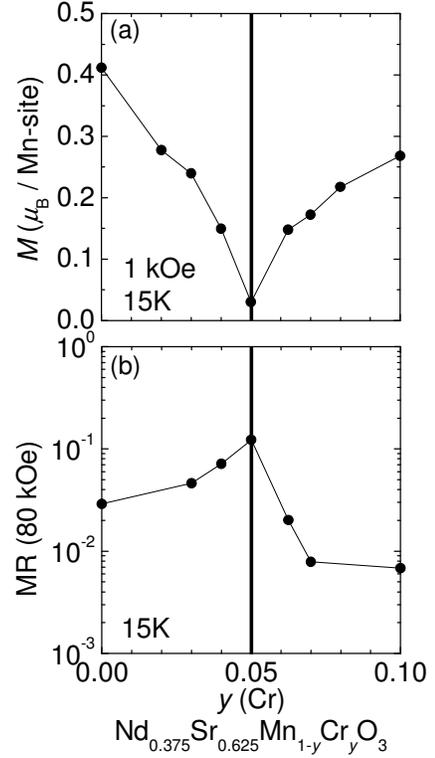}
\caption{\label{fig:fig3} Cr-doping concentration $y$ dependence of (a) $M$ under $H$=1 kOe and (b) MR(80 kOe) at 15 K.}
\end{figure}
  We plot the $M$ under $H$ = 1 kOe and MR(80 kOe) at 15K as a function of $y$ in Figs. 3(a) and 3(b), respectively. These figures, as mentioned above, demonstrate that the F correlation and MR are most suppressed at $y$ = 0.05 and show the reentrant behavior, implying that a phase boundary is located at $y = 0.05$, or a phase crossover occurs across $y = 0.05$.

 Let us discuss the origin of the reentrant behavior. In $0 \leq y \leq 0.05$, the F correlation and MR are systematically reduced with increasing $y$. This behavior is quite similar to that observed in NSMO with $x = 0.625$, the F correlation and MR of which are also suppressed with increasing hole-concentration $x$ from 0.625\cite{Akahoshi_PRB_77}. Therefore, we interpret that Cr- and hole-doping have almost the same effect on the magnetic and transport properties of NSMO with $x = 0.625$ in $0 \leq y \leq 0.05$. This is probably because Cr$^{3+}$ and Mn$^{4+}$ have the same electronic configuration of $t_{2g}^3e_{g}^0$. In $y > 0.05$, the F correlation and MR reappear; the F correlation is developing with further increasing $y$ from 0.05. This reminds us the fact that Cr-doping into CO/OO Nd$_{0.5}$Ca$_{0.5}$MnO$_3$ produces the FM clusters embedded in the CO/OO matrix\cite{Kimura_PRB_62,Kimura_PRL_83}. In NSMO with $x = 0.625$ as well as Nd$_{0.5}$Ca$_{0.5}$MnO$_3$, the F correlation is probably induced around Cr$^{3+}$. However, the F correlation is not so strong compared with that of Cr-doped Nd$_{0.5}$Ca$_{0.5}$MnO$_3$, the reason for which might be explained by the fact that NSMO with $x = 0.625$ is apart from the FM state, which is often found in the low-doped ($x \leq 0.5$) perovskite manganites. The reentrant behavior observed in Cr-doped NSMO with $x = 0.625$ is perhaps due to competition between the hole-doping effect and the FM cluster effect caused by Cr-doping. In $0 \leq y \leq 0.05$, the number (or the size) of the FM clusters is so small that the hole-doping effect is dominant. With increasing $y$, the number (or the size) of the FM clusters is becoming large, and the FM cluster effect finally overcomes the hole-doping effect in $y > 0.05$. As a result, the F correlation is macroscopically observed again in the magnetic and transport properties of NSMCO with $y > 0.05$. The detailed mechanism of the reentrant behavior is now under investigation.

%
%

%
%
\subsection{\label{sec:Acknowledgment}Acknowledgment}

 We thank Y. Izuchi for her help in growing single crystals and the measurements using PPMS. This work was partly supported by the Mazda Foundation, the Asahi Glass Foundation, and Grant-in-Aid for scientific research (C) from the Japan Society for Promotion of Science.

\end{document}